\documentstyle[aps,12pt]{revtex}

\begin{document}
\author{J.P. Krisch and E.N. Glass\thanks{%
Permanent address: Physics Department, University of Windsor, Ontario N9B
3P4, Canada}}
\address{Department of Physics, University of Michigan, Ann Arbor, Michgan 48109}
\date{12 February 2001, \ \ \ preprint MCTP-01-08}
\title{Critical Exponents For Schwarzschild-Kerr and BTZ Systems}
\maketitle

\begin{abstract}
\newline
\newline
Regarding the spin-up of Schwarzschild-Kerr and
Ba\~{n}ados-Teitelboim-Zanelli systems as a symmetry breaking phase
transition, critical exponents are evaluated and compared with classical
Landau predictions. We suggest a definition of isothermal compressibity
which is independent of spin direction.\ We find identical exponents for
both systems, and possible universality in the phase transitions of these
systems.\newline
\newpage\ 
\end{abstract}

\section{INTRODUCTION}

Black holes are perfect absorbers classically but do not emit anything;
their classical physical temperature is absolute zero. In the semi-classical
approximation black holes emit Hawking radiation with a perfect thermal
spectrum. This allows a consistent interpretation of the laws of black hole
mechanics as physically corresponding to the ordinary laws of
thermodynamics. The classical laws of black hole mechanics together with the
formula for the temperature of Hawking radiation allow one to identify the
general relativistic horizon area, ${\cal A}/4$, as playing the mathematical
role of entropy. The geometric thermodynamics of black holes has provided
many useful insights to the relation between geometric quantities and
physics. These analogies are particularly useful when applied to constant
mass black hole metric transformations involving a symmetry reduction. The
Schwarzschild to Kerr (SCH-K) transform is an example. In thermodynamics,
these transitions involve order parameters and critical exponents linked to
response functions (such as specific heat or magnetic susceptibility)
describing phase transitions. Existing thermodynamic analogies \cite{Dav77}, 
\cite{Dav78} suggest that the SCH-K transform can be viewed as a phase
transition. The Schwarzschild solution is spherically symmetric and has
higher symmetry than axially symmetric Kerr. We interpret the SCH-K
transition, wherein Schwarzschild ''spins up'', as symmetry reduction, and
thus assign $J$ the role of order parameter.

Three elements of critical phenomena, power law behavior, universality, and
scaling, have become increasingly important in general relativity and have
been identified in some general relativistic systems. Davies \cite{Dav77}
and Lousto \cite{Lou95} investigated the possibility of a phase transition
in the SCH-K system at $J\simeq .68m^{2}$, a point where the specific heat
has an infinite discontinuity. The identification of this point with a phase
transition has been further discussed by Katz et al \cite{KOK93a},\cite
{KOK93b}, and Kaburaki \cite{Kab94},\cite{Kab96}. The three elements have
recently been shown to single out a critical solution for some general
relativistic solution sets. In 2+1 dimensions Cai, Lu, and Zhang \cite{CLZ97}
have considered the transition to extremal black hole solutions as a
critical transition and discussed this behavior in the 2+1 solution sets.
Cai et al point out that Curir \cite{Cur81a},\cite{Cur81b} claims a critical
point exists at the extremal limit of the hole and a phase transition occurs
from extremal to non-extremal Kerr black hole. She has also investigated the
spin interaction of the inner and outer Kerr horizons \cite{Cur86},\cite
{Cur89}.

In 3+1 dimensions Choptuik \cite{Cho93} discovered universal power law
scaling of black hole mass for critical data in the gravitational collapse
of a minimally coupled scalar field. Critical collapse for other matter
distributions is reviewed by Gundlach \cite{Gun99a}, with an extensive list
of references. Including angular momentum in families of collapsing matter
solutions allows the study of angular momentum behavior at the black hole
threshold in critical collapse. For fluids with equation of state $p=k\rho $%
, the angular momentum vector has a critical exponent proportional to the
mass critical exponent found by Choptuik.

In this paper we evaluate the critical exponents for $J=0\rightarrow J\neq 0$
transition in two systems: the 3+1 asymptotically flat SCH-K system and the
2+1 Ba\~{n}ados-Teitelboim-Zanelli (BTZ) black hole. In each case we take
the Hawking temperature of the $J=0$ horizon as the critical temperature.
Surprisingly, the results are the same for both systems and are also similar
to those derived from the classical Landau theory and mean field theory \cite
{Sta71}. This is a different approach to studying the effects of adding
angular momentum than during the formation of the black hole solutions in
critical collapse. The black holes we consider already exist, and angular
momentum is added. We do not study the role of angular momentum in the
collapse process. Even so, there are some similarities between critical
collapse and the critical behavior we find in the black hole spin-up process.

In the next two sections we briefly review the SCH-K thermodynamics needed
to write the response functions and examine critical behavior. The BTZ
solution and its critical exponents are discussed in the fourth section.
Rotational scaling inequalities are derived in section V. In the final
section we discuss the similarity to classical Landau theory and
universality, and the relation to critical collapse phenomena.

\section{SCH-K THERMODYNAMICS}

The Kerr solution, characterized by mass '$m$' and specific angular momentum
'$a$'$=J/m$, is given in Boyer-Lindquist $(t,r,\theta ,\varphi )$
coordinates by 
\begin{eqnarray}
ds^{2} &=&\Psi dt^{2}-(\Sigma /\Delta )dr^{2}-\Sigma (d\theta ^{2}+\sin
^{2}\theta \ d\varphi ^{2}) \\
&&+(1-\Psi )a\text{sin}^{2}\theta \ 2dtd\varphi -(2-\Psi )a^{2}\sin
^{4}\theta \ d\varphi ^{2},  \nonumber
\end{eqnarray}
where $\Sigma =r^{2}+a^{2}\cos ^{2}\theta $, $\Psi =1-2mr/\Sigma $, $\Delta
=r^{2}-2mr+a^{2}$. When $a=0$, $\Psi =1-2m/r$ and the Schwarzschild solution
is recovered.

It is well known that the SCH-K family of metrics is conformally scaled by
the mass parameter such that 
\[
g_{\mu \nu }^{Kerr}(t,r,m,a)=(m/M)^{2}g_{\mu \nu }^{Kerr}(T,R,M,A) 
\]
where $T/M=t/m$, $R/M=r/m$, $A/M=a/m$. We will consider a thermodynamic
system of SCH-K metrics with energies labelled by '$m$', and focus our
attention on the horizons of this system, i.e. the ''black hole'' states 
\cite{Wal84},\cite{Gou00}. The thermodynamic parameters are $%
m,T_{_{H}},S_{_{BH}},\Omega _{h},J.$

The horizons are 2-surfaces of constant $t$ and $r$. The outer one is
located at $\Delta (r_{+})=0$ with $r_{+}=m+\sqrt{m^{2}-a^{2}}$. The horizon
area is 
\begin{equation}
{\cal A}=\oint\limits_{r=r_{+}}\sqrt{g_{\theta \theta }g_{\varphi \varphi }}%
\ d\theta d\varphi =4\pi (r_{+}^{2}+a^{2}).
\end{equation}
The angular velocity of the outer horizon, $\Omega _{h}:=(\frac{d\varphi }{dt%
})_{r_{+}}$, is given by 
\begin{equation}
\Omega _{h}=\frac{a}{r_{+}^{2}+a^{2}}.  \label{ang-vel}
\end{equation}

The Bekenstein-Hawking entropy is $S_{_{BH}}=(\frac{k_{B}c^{3}}{\hbar G})%
{\cal A}/4={\cal A}/4$ in units with $k_{B}=\hbar =c=G=1$. 
\begin{equation}
S_{_{BH}}=\pi (r_{+}^{2}+a^{2})=2\pi m^{2}[1+\sqrt{1-J^{2}/m^{4}}].
\label{bh-ent}
\end{equation}
Hawking radiation is the energy flux from the black hole and considered as
black body radiation in a thermal bath at temperature $T_{_{H}}$. Using
angular momentum as an extensive thermodynamic parameter, the temperature of
the black hole at constant $J$ provides the relations: 
\begin{mathletters}
\begin{eqnarray}
T_{_{H}} &=&\left[ \frac{\partial m\,\text{\ }\ }{\partial S_{_{BH}}}\right]
_{J}  \label{th-tempa} \\
&=&\frac{1}{4\pi m}\left[ \frac{(1-J^{2}/m^{4})^{1/2}}{%
1+(1-J^{2}/m^{4})^{1/2}}\right]  \label{th-tempb} \\
&=&\frac{1}{8\pi m}[1-(2\pi J/S_{_{BH}})^{2}].  \label{th-tempc}
\end{eqnarray}
$T_{_{Sch}}=1/(8\pi m)$ is the Hawking temperature of the Schwarzschild
horizon \cite{Wal94}.

The rotational analog of $p=p(V,T)$ is $\Omega _{h}=\Omega _{h}(J,T_{_{H}})$
or $J=J(\Omega _{h},T_{_{H}})$. To find the equation of state we rewrite Eq.(%
\ref{ang-vel}) as 
\end{mathletters}
\begin{equation}
m\Omega _{h}=\frac{J/m^{2}}{2[1+\sqrt{1-(J/m^{2})^{2}}]}.  \label{ang-vel2}
\end{equation}
Dividing Eq.(\ref{ang-vel}) by Eq.(\ref{th-tempb}) yields 
\[
\phi =\frac{J/m^{2}}{\sqrt{1-(J/m^{2})^{2}}},\text{\ \ \ }\phi :=\Omega
_{h}/(2\pi T_{_{H}}) 
\]
with inversion 
\begin{equation}
J=m^{2}\frac{\phi }{(1+\phi ^{2})^{1/2}}.  \label{j-phi}
\end{equation}
Using Eq.(\ref{th-tempb}) to eliminate $m^{2}$, we obtain the equation of
state for Kerr black holes: 
\begin{equation}
J=\frac{\phi }{4(2\pi T_{_{H}})^{2}}(1+\phi ^{2})^{-1/2}\,[1+(1+\phi
^{2})^{1/2}]^{-2}.  \label{j-omega-t}
\end{equation}
These thermodynamic functions will be used to generate the response
functions which, in turn, lead to the critical exponents.

\section{SCH-K CRITICAL BEHAVIOR}

\subsection{Response Functions}

Two of the systems where critical behavior has been studied are fluid/gas
and magnetic systems. We will examine response functions: specific heat,
compressibility, susceptibility. Critical exponents are defined with the use
of a dimensionless temperature deviation $\epsilon :=(T-T_{c})/T_{c}$ where $%
T_{c}$ is the critical temperature. The most common critical exponents \cite
{Yeo94} are

\subsubsection{Fluid/Gas Systems}

\[
\begin{tabular}{ll}
specific heat at constant volume & $\ \ \ C_{V}\approx \mid \epsilon \mid
^{-\alpha }$ \\ 
fluid/gas density difference & $\ \ \ \rho _{fl}-\rho _{gas}\approx
(-\epsilon )^{\beta }$ \\ 
isothermal compressibility & $\ \ \ \kappa _{\text{{\sc T}}}\approx \mid
\epsilon \mid ^{-\gamma }$ \\ 
critical isotherm ($\epsilon =0$) & \ \ \ $p-p_{c}\approx \mid \rho
_{fl}-\rho _{gas}\mid ^{\delta }$ sgn($\rho _{fl}-\rho _{gas}$).
\end{tabular}
\]

\subsubsection{Magnetic Systems}

\[
\begin{tabular}{ll}
zero-field specific heat & $\ \ \ C_{_{H}}\approx \mid \epsilon \mid
^{-\alpha }$ \\ 
zero-field magnetization & $\ \ \ {\cal M}\approx \mid \epsilon \mid ^{\beta
}$ \\ 
zero-field isothermal susceptibility & $\ \ \ \chi _{_{T}}\approx \mid
\epsilon \mid ^{-\gamma }$ \\ 
critical isotherm ($\epsilon =0$) & \ \ \ $H\approx \mid {\cal M}\mid
^{\delta }$ sgn($M$).
\end{tabular}
\]

\subsection{Rotating Systems}

Rotating systems display thermodynamic similarity to magnetic systems. The
angular momentum of the Kerr system provides an interesting analogy. Smarr's
formula \cite{Sma73} 
\begin{eqnarray}
m &=&2[(\kappa /2\pi )({\cal A}/4)+\Omega _{h}J]  \label{smarr1} \\
&=&2[T_{_{H}}S_{_{BH}}+\Omega _{h}J]  \nonumber
\end{eqnarray}
is obtained here using Eqs.(\ref{ang-vel}), and (\ref{bh-ent}). Equation (%
\ref{smarr1}) is analogous to the Euler relation $U=TS-{\cal M}H$, where $%
{\cal M}H$ is magnetic energy for magnetization ${\cal M}$. The analogy is
more striking if one writes the magnetic energy in quantum form as $-g\mu
_{_{B}}HJ$, with $g$ the Land\'{e} g-factor, $\mu _{_{B}}$ the Bohr
magneton, and $J=m\hbar $ (here $m$ is the magnetic quantum number). Just as 
${\cal M}$ and $H$ form an extensive/intensive pair, so do $J$ and $\Omega
_{h}$.

Critical exponents for rotating systems are 
\begin{equation}
\begin{tabular}{ll}
specific heat & $\ \ \ C_{\Omega }\approx \mid \epsilon \mid ^{-\alpha }$ \\ 
angular momentum & $\ \ \ J\approx \mid \epsilon \mid ^{\beta }$ \\ 
isothermal compressibility & $\ \ \ \kappa _{\text{{\sc T}}}\approx \mid
\epsilon \mid ^{-\gamma }$ \\ 
critical isotherm ($\epsilon =0$) & \ \ \ $\Omega _{h}\approx \mid J\mid
^{\delta }$ sgn($J$).
\end{tabular}
\label{rot-exps}
\end{equation}
For critical temperature $T_{c}=T_{_{Sch}}$, the temperature deviation is 
\begin{equation}
\epsilon =\frac{T_{_{H}}-T_{c}}{T_{c}}=\frac{(1-J^{2}/m^{4})^{1/2}-1}{%
(1-J^{2}/m^{4})^{1/2}+1},  \label{eps}
\end{equation}
with the inversion 
\begin{equation}
(a/m)^{2}=-\frac{4\epsilon }{(1-\epsilon )^{2}},\text{\ \ \ }a\Omega _{h}=%
\frac{\epsilon }{\epsilon -1}.  \label{a-eps}
\end{equation}

\subsection{Response Functions for SCH-K}

\subsubsection{Specific Heat}

Given the entropy and temperature in Eq.(\ref{th-tempc}), the specific heat
at constant $J$ is 
\begin{equation}
C_{J}=\left[ T_{_{H}}\frac{\partial S_{_{BH}}}{\partial T_{_{H}}\ }\right]
_{J}=\frac{4\pi mT_{_{H}}S_{_{BH}}}{1-8\pi mT_{_{H}}-4\pi
T_{_{H}}^{2}S_{_{BH}}}.  \label{cj}
\end{equation}
Equation (\ref{cj}) agrees with the form given by Davies \cite{Dav77} upon
using $T_{_{H}}\rightarrow T_{_{H}}/8\pi $ and $S_{_{BH}}\rightarrow 8\pi
S_{_{BH}}$. In the Schwarzschild limit $C_{J=0}=-8\pi m=-m/T_{_{Sch}}$,
where the negative sign indicates the well known idea that Schwarzschild
black holes get hotter as they lose mass.

We write Eq.(\ref{cj}) as 
\begin{equation}
C_{J}=-\frac{4\pi m^{2}J(J-2\Omega _{h}m^{3})}{J^{2}-12\Omega _{h}^{2}m^{6}}.
\label{cj-omeg2}
\end{equation}
A discontinuity in $C_{J}$ occurs at $J_{\infty }=\sqrt{12}\Omega _{h}m^{3}$%
, or $a_{\infty }=2\sqrt{3}\Omega _{h}m^{2}$. With $\Omega _{h}$ expressed
in Eq.(\ref{ang-vel}) we have 
\begin{equation}
a_{\infty }=\frac{\sqrt{3}a_{\infty }}{1+\sqrt{1-(a_{\infty }/m)^{2}}}.
\label{a-discont}
\end{equation}
For $a_{\infty }=xm$, $x^{2}=2\sqrt{3}-1$, or $a_{\infty }\simeq .68m$. The
temperature of the discontinuity is 
\begin{equation}
4\pi T_{_{H}}(a_{\infty })\simeq .42/m.  \label{temp-discont}
\end{equation}
Lousto \cite{Lou95} has used this temperature as the critical temperature
for deriving critical exponents. The presence of this discontinuity in $%
C_{J} $ means that the Helmholtz free energy (as well as the Gibbs
potential) does not have a valid Taylor expansion \cite{Sta71} in term of $J$%
. Such an expansion must exist in order for Landau theory to predict
critical exponents.

In the following, we use $4\pi T_{_{Sch}}=.5/m$ as the critical temperature
since symmetry reduction occurs at the $J=0$ to $J\neq 0$ transition. There
is no discontinuity in the specific heat at $T_{_{Sch}}$.

The specific heat at constant $\Omega _{h}$ is 
\begin{eqnarray}
C_{\Omega } &=&-\frac{16\pi mT_{_{H}}S_{_{BH}}}{(1+4\Omega _{h}^{2}m^{2})^{2}%
}  \label{c-omega} \\
&=&-\frac{\pi J^{2}(J-2\Omega _{h}m^{3})}{4\Omega _{h}^{3}m^{7}}.  \nonumber
\end{eqnarray}

\subsubsection{Isothermal Compressibility}

Davies \cite{Dav77} indicates that $C_{\Omega }-C_{J}$ measures how the
black hole can be spun up at constant temperature, with the difference in
specific heats depending on $\kappa _{\text{{\sc T}}}$. The form suggested
by Davies is spin direction dependent: 
\begin{mathletters}
\begin{eqnarray}
\kappa _{\text{{\sc T}}}^{{\small Dav}} &=&-\frac{1}{J}\left[ \frac{\partial
J}{\partial \Omega _{h}}\right] _{T_{_{H}}}  \label{kdav-a} \\
&=&-(1/\Omega _{h})+8(J/m)-4\Omega _{h}(J/m)^{2}.  \label{kdav-b}
\end{eqnarray}
Since $\Omega _{h}(-a)=-\Omega _{h}$, and $J(-a)=-J$, it follows that $%
\kappa _{\text{{\sc T}}}^{Dav}(-a)=-\kappa _{\text{{\sc T}}}^{Dav}$. There
are other possibilities. We could follow the definition of magnetic
susceptibility $\chi _{_{T}}$ as the fluid/gas analogy of isothermal
compressibility including no powers of $J$ or we could include even powers
of $n$ where 
\end{mathletters}
\begin{equation}
\kappa _{\text{{\sc T}}}^{rot(n)}=-\frac{1}{J^{n}}\left[ \frac{\partial J}{%
\partial \Omega _{h}}\right] _{T_{_{H}}}.  \label{k-rotn}
\end{equation}
Clearly $\kappa _{\text{{\sc T}}}^{rot(n)}(-a)=\kappa _{\text{{\sc T}}%
}^{rot(n)}$ for $n$ even.

\subsection{Critical Exponents for SCH-K}

The critical exponents to be found are $\alpha $ from the specific heat, $%
\beta $ from the angular momentum, $\gamma $ from the compressibility, and $%
\delta $ from the critical isotherm.

\paragraph{\ }

$\alpha =0$ from the expansion of $C_{\Omega }$ in Eq.(\ref{c-omega}). $%
C_{\Omega }=-8\pi m^{2}(1+4\epsilon +\ldots )$. The value is unchanged if we
expand $C_{J}$ in Eq.(\ref{cj}) in analogy with $C_{V}$ since $C_{J}=-8\pi
m^{2}(1-5\epsilon +\ldots )$.\newline

\paragraph{\ }

$\beta =1/2$ from the inversion of Eq.(\ref{eps}). $J=2m^{2}\mid \frac{%
\epsilon ^{1/2}}{1-\epsilon }\mid $.\newline

\paragraph{\ }

$\gamma =n/2=0,$ $1/2$, or $1$. The value $1/2$ is computed from Davies'
formula (\ref{kdav-b}) 
\begin{eqnarray*}
\frac{1}{\kappa _{\text{{\sc T}}}^{Dav}} &=&-\frac{\Omega _{h}}{1-8a\Omega
_{h}+4(a\Omega _{h})^{2}} \\
&\simeq &\frac{\mid \epsilon \mid ^{1/2}}{2m(1+7\epsilon +4\epsilon ^{2})}.
\end{eqnarray*}
The other values follow from $n=0$ and $n=2$ in Eq.(\ref{k-rotn}).\newline

\paragraph{\ }

$\delta =1$. The critical isotherm is given by $\Omega _{h}=J(1-\epsilon
)/(4m^{3})$.

\section{CRITICAL EXPONENTS AND THE BTZ SOLUTION}

One of the motives for finding critical exponents is the possibility that
their validity will extend beyond the single system from which they were
obtained and thus be somewhat ''universal''. We have calculated the
exponents for the SCH-K horizons. Another system with horizon thermodynamics
for which critical exponents can be calculated is the anti-de Sitter (AdS)
family of black holes of varying dimensions. They differ from the SCH-K
system in several respects. The AdS solutions are not asymptotically flat,
there is no unique asymptotic timelike Killing norm, and they have a
cosmological constant. As a first check of possible universality, we
calculate the critical exponents for the 2+1 BTZ black hole.

The Lorentzian 2+1 section of the BTZ black hole \cite{HHT-R99} is 
\begin{equation}
ds^{2}=N^{2}dT^{2}-\rho ^{2}[d\Phi -(\frac{j}{2\rho ^{2}})dT]^{2}-(y/\rho
)^{2}N^{-2}dy^{2}.  \label{btz-met}
\end{equation}
$\Phi $ is $2\pi $ periodic. The outer horizon has radius $y_{+}$ where 
\begin{mathletters}
\begin{eqnarray}
y_{+}^{2} &=&(m/l^{2})[1-(jl/m)^{2}]^{1/2},  \label{btz-val1} \\
2\rho ^{2}(y) &=&2y^{2}-y_{+}^{2}+m/l^{2},  \label{btz-val2} \\
N^{2}(y) &=&(yl/\rho )^{2}(y^{2}-y_{+}^{2}).  \label{btz-val3}
\end{eqnarray}
The AdS solution as a bound state of BTZ is fixed by $m=-1$ and $j=0$. This
black hole has temperature $T$, entropy $S$, and horizon rotation speed $%
\Omega $ just as the Kerr system does. For the AdS black hole \cite{HHT-R99}
we have 
\end{mathletters}
\begin{equation}
T=\frac{l\sqrt{2m}}{2\pi }\left[ \frac{1-(jl/m)^{2}}{1+\sqrt{1-(jl/m)^{2}}}%
\right] ^{1/2}.  \label{bzt-temp}
\end{equation}
\begin{equation}
S=\frac{\pi }{2}\rho (y_{+})  \label{bzt-ent}
\end{equation}
\begin{equation}
\Omega =-\frac{j}{2\rho ^{2}(y_{+})}  \label{btz-omega}
\end{equation}
We take the critical temperature for the spin-up transition as the
temperature of the non-rotating hole 
\[
T_{c}=\frac{l\sqrt{m}}{2\pi }. 
\]
The temperature deviation for the critical parameters is 
\begin{equation}
\epsilon =\frac{T-T_{c}}{T_{c}}=\sqrt{2}\left[ \frac{1-(jl/m)^{2}}{1+\sqrt{%
1-(jl/m)^{2}}}\right] ^{1/2}-1,  \label{btz-eps}
\end{equation}
with the inversion 
\[
(jl/m)^{2}=1-\frac{(1+\epsilon )^{2}}{16}[1+\epsilon +(9+2\epsilon +\epsilon
^{2})^{1/2}]^{2}. 
\]
The $\beta $ exponent follows directly from this expansion and we find $%
\beta =1/2.$ From Eq.(\ref{btz-omega}) we have 
\begin{equation}
\Omega \simeq -\frac{j}{m/l^{2}+\sqrt{m/l^{2}}}
\end{equation}
so that $\delta =1.$ $\alpha =0$ follows from the specific heat 
\begin{equation}
C_{\Omega }=\left[ T\frac{\partial S}{\partial T}\right] _{\Omega }\simeq 
\sqrt{\frac{m\pi ^{2}}{8l^{2}}}.
\end{equation}
Just as in the SCH-K calculation, $\left[ \frac{\partial \Omega }{\partial J}%
\right] _{T}$ is constant as one approaches the critical point. The critical
behavior is determined by the power of $J$ in the compressibility and we
find $\gamma =n/2$. Evaluating the thermodynamic functions shows the
critical exponents to be identical to those calculated for the 3+1 SCH-K
system.

\section{SCALING INEQUALITIES}

The critical exponents can be constrained by inequalities based on
thermodynamic arguments. Two of the most useful \cite{Sta71} are the
Rushbrooke inequality 
\[
\alpha +2\beta +\gamma \geq 2 
\]
and Widom's (actually Griffiths' \cite{Sta71}) 
\[
\gamma \geq \beta (\delta -1). 
\]
The values predicted by the classical theory of Landau \cite{LL58} are 
\begin{equation}
\alpha =0,\text{ }\beta =1/2,\text{ }\gamma =1,\text{ }\delta =3,
\label{params1}
\end{equation}
and they satisfy both inequalities. These inequalities need to be examined
for rotating systems.

To prove the Rushbrooke inequality, consider the relation between the
specific heats at constant angular momentum and constant angular velocity: 
\begin{equation}
C_{\Omega }=C_{J}+\frac{TJ^{n}}{2J^{2}\kappa _{\text{{\sc T}}}^{rot(n)}}%
\left( \left[ \frac{\partial J}{\partial T}\right] _{\Omega _{h}}\right)
^{2}.  \label{co-cj}
\end{equation}
The specific heats are negative in the region around $J=0$. Explicitly
including the negative signs, (\ref{co-cj}) can be rewritten as 
\[
\mid C_{\Omega }\mid =\mid C_{J}\mid +\frac{TJ^{n}}{2J^{2}\mid \kappa _{%
\text{{\sc T}}}^{rot(n)}\mid }\left( \left[ \frac{\partial J}{\partial T}%
\right] _{\Omega _{h}}\right) ^{2} 
\]
or 
\[
\mid C_{\Omega }\mid \ >\frac{TJ^{n}}{2J^{2}\mid \kappa _{\text{{\sc T}}%
}^{rot(n)}\mid }\left( \left[ \frac{\partial J}{\partial T}\right] _{\Omega
_{h}}\right) ^{2}. 
\]
Substituting the critical behavior, we find the Rushbrooke rule for rotating
systems 
\begin{equation}
\alpha +n\beta +\gamma -2\geq 0.  \label{proof-rush}
\end{equation}

Proof of the Widom-Griffiths inequality starts with $\Omega \simeq J^{\delta
}$ and $J\simeq \epsilon ^{\beta }$ from Eq.(\ref{rot-exps}). Thus 
\[
\frac{\partial \Omega }{\partial J}\simeq \delta J^{\delta -1}\simeq \delta
\epsilon ^{\beta (\delta -1)}. 
\]
Eqs. (\ref{rot-exps}) and (\ref{k-rotn}) express 
\[
\frac{1}{\kappa _{\text{{\sc T}}}^{rot(n)}}=-J^{n}\left[ \frac{\partial
\Omega }{\partial J}\right] \simeq \epsilon ^{\gamma } 
\]
or 
\[
\frac{\partial \Omega }{\partial J}\simeq J^{-n}\epsilon ^{\gamma }\simeq
\epsilon ^{\gamma -n\beta }. 
\]
It follows that $\epsilon ^{\gamma +\beta (1-\delta -n)}\simeq \delta $. As
one approaches the critical point and $\epsilon \rightarrow 0$, the previous
expression can be constant only if the exponent is zero: $\gamma =\beta
(\delta -1)+n\beta $. Since $n$ is defined to range over positive even
integers and $\beta $ must be positive, the inequality is expressed as 
\begin{equation}
\gamma \geq \beta (\delta -1).  \label{proof-wid}
\end{equation}

The four exponent values computed here are $\alpha =0$, $\beta =1/2$, $%
\gamma =0$, $1/2$, or $1$, $\delta =1$. They satisfy the Widom-Griffiths
inequality for all values. Rushbrooke's is satisfied for $\gamma =1$ and $%
n=2 $.

\section{DISCUSSION}

We have investigated the critical behavior of the spin-up transition in the
3+1 Schwarzschild-Kerr system and the 2+1 BTZ system. One of the striking
characteristics of critical phenomena is the fact that many measures of a
system's behavior near a critical point are quite independent of the details
of the interactions between the particles making up the system. The
universal features are not only independent of the numerical details of the
interparticle interactions, but are also independent of the most fundamental
aspects of the structure of the system. The critical exponents for the two
systems we examined are the same. This result is interesting because of the
differences in the physical dimension and geometries of the two spacetimes.
One should note that the dimension of the thermodynamic space has not
changed so long as the cosmological constant is not a thermodynamic
variable. The agreement of the two sets of critical exponents for the
spin-up transition indicates universality may exist for the $J=0\rightarrow
J\neq 0$ transition but the other higher dimensional AdS spaces need to be
checked as well as the effect of increasing the dimension of the
thermodynamic space by considering a thermodynamic cosmological constant.

Another interesting aspect of the critical exponents calculated in this
paper is their similarity to the classic Landau exponents, which are $\alpha
=0,$ $\beta =1/2,$ $\gamma =1,$ $\delta =3$.

The calculation of these numbers is based on a series expansion of the
Helmholtz potential $F=m-TS$ as a function of order parameter $J$: 
\begin{equation}
F(J,T)=L_{0}(T)+L_{2}(T)J^{2}+\ldots  \label{f-j-t-expan}
\end{equation}
where the coefficients can be expanded about the critical temperature 
\begin{equation}
L_{i}(T)=l_{i(0)}+l_{i(1)}(T-T_{c})+\ldots  \label{l-coef}
\end{equation}
Some of the thermodynamic response functions examined for critical behavior
are simple derivatives of the Helmholtz free energy. Following Stanley \cite
{Sta71} we write 
\begin{equation}
\Omega =\left[ \frac{\partial F}{\partial T}\right]
_{J}=2L_{2}(T)J+4L_{4}(T)J^{3}+\ldots  \label{om-f}
\end{equation}
\begin{equation}
\left[ \frac{\partial \Omega }{\partial J}\right] _{T}=\frac{\partial ^{2}F}{%
\partial T^{2}}=2L_{2}(T)+12L_{4}(T)J^{2}+\ldots  \label{dom-dj-f}
\end{equation}
From the expression for $\Omega $ we see that if $l_{2(0)}$ is non-zero then
the isothermal derivative of the horizon angular speed is constant at the
critical point. This behavior agrees with the direct calculations done in
the previous sections producing $\delta =1$ and a $\gamma $ that reflects
the power of $J$ multiplying the derivative in the compressibility
definition (\ref{k-rotn}). In the Landau model for the magnetic critical
exponent, $l_{2(0)}$ is set to zero based on expecting the inverse zero
field susceptibility to be zero. The magnetic critical $\gamma $ follows
directly from the expansion with $l_{2(0)}=0$ and $l_{2(1)}$ providing the $%
\gamma =1$ behavior with no multiplying $J$ factor in the susceptibility
definition. With $l_{2(0)}=0$, the Landau model predicts $\delta =3$. If, in
these magnetic expansions, $l_{2(0)}$ is not set to zero, then the Helmholtz
expansion produces the same $\gamma $ and $\delta $ critical exponents that
are found directly for the rotating systems.

The other critical exponents, $\alpha $ and $\beta $, agree for both the
Landau model and the rotational calculation. The $\alpha =0$ exponent also
follows from the Helmholtz expansion with $l_{2(0)}$ non-zero, even though
in the magnetic case a more complicated argument is needed to produce this
behavior. The magnetic $\beta =1/2$ is the single exponent that does not
follow from the expansion but from the definition of the black hole
temperature. In the Landau model, it is calculated by looking at the zero
field magnetization (with $J$ analogous to ${\cal M}$) in Eq.(\ref{om-f}).
There is no non-zero $J$ for zero $\Omega $ in the rotational system when $J$
is assumed to come from purely kinematic rotation.

As pointed out in the introduction, investigating angular momentum during
critical collapse studies the effects of initial data sets which break
spherical symmetry during collapse, while here we have studied the effects
of breaking spherical symmetry by adding angular momentum to already
existing black hole solutions. We found from the inversion of 
\[
\epsilon =\frac{(1-J^{2}/m^{4})^{1/2}-1}{(1-J^{2}/m^{4})^{1/2}+1} 
\]
that 
\[
J=2m^{2}\mid \frac{\epsilon ^{1/2}}{1-\epsilon }\mid . 
\]
Choptuik \cite{Cho93} studied the spherically symmetric collapse of a
massless scalar field and found that the mass scaled as 
\[
m\approx (p-p_{\ast })^{\gamma } 
\]
where $p$ indexes a one-parameter set of initial data and $\gamma \simeq
0.37.$ Garfinkle et al \cite{GGM-G99} investigated the effects of breaking
spherical symmetry with angular momentum during the gravitational collapse
of a massless scalar field, and found that the angular momentum scaled as 
\[
J\approx (p-p_{\ast })^{\mu } 
\]
with $\mu \approx 0.76$ or $\mu \sim 2\gamma $, and where $\gamma $ depends
on the type of collapsing matter$.$ This suggests that the pure
thermodynamic relation between $J$ and $m$ may be reflected in critical
collapse behavior. We note that the thermodynamic system is an equilibrium
system and the other is not. Gundlach \cite{Gun99b} investigated the
critical collapse of a perfect fluid with equation of state $p=k\rho $ and
found that the angular momentum critical exponent was 
\[
\mu =\gamma \frac{5(1+3k)}{3(1+k)} 
\]
for mass exponent $\gamma $ and a range of $k$. This is not reflected in the
mass dependence of the thermodynamic relation. Gundlach has emphasized three
final endpoints of collapse: a black hole, a star, or a dispersive explosion
with no remnant. Perfect fluid initial data sets seem too restrictive to
model all of Gundlach's possibilities. Perfect fluids have not been found as
interiors for the Kerr vacuum. The mass-angular momentum relation in
thermodynamics and in critical collapse, and the relevance of perfect fluid
collapse to non-black hole states should be interesting to investigate.

In conclusion, we have extended classic scaling inequalities to include
rotation, and have defined an isothermal compressibity which is independent
of spin direction. We have calculated four of the critical exponents for the
symmetry breaking spin-up of both a 3+1 and 2+1 black hole. The critical
exponents for the two cases agree with each other and with some of the
values calculated from a Landau type Helmholtz expansion. The agreement may
indicate a universality in spin-up behavior.

\end{document}